\documentstyle[12pt,epsf]{article}

\def\ru1{\rule[-0.4truecm]{0mm}{1truecm}}

\def\be{\begin{eqnarray}}
\def\ee{\end{eqnarray}}
\def\bea{\begin{eqnarray}}
\def\eea{\end{eqnarray}}
\def\kT{{\bf k}_\perp}
\def\rT{{\bf r}_\perp}

\def\qT{{\bf q}_\perp}

\def\bT{{\bf b}_\perp}
\def\cT{{\bf c}_\perp}
\def\IT{{\bf I}_\perp}

\def\DT{{\bf \Delta}_\perp}
\def\D2{{\bf \Delta}_\perp^2}
\def\0T{{\bf 0}_\perp}
\begin{document}
\title{Sivers Asymmetry and Generalized Parton Distributions in
Impact Parameter Space}
\author{Matthias Burkardt\\
Department of Physics, New Mexico State University,\\
Las Cruces, NM 88003-0001, U.S.A.\\[4.ex]
Dae Sung Hwang\\
Department of Physics, Sejong University,\\ 
Seoul 143-747, Korea}

\date{\today}% It is always \today, today,
\maketitle
\begin{abstract}
Recently, it has been pointed out that there exists a connection 
between 
the generalized parton distribution $E(x,0,t)$ and the Sivers 
asymmetry. For transversely polarized nucleon targets, generalized 
parton distributions are asymmetric in impact parameter space. 
This impact parameter space  asymmetry, together with the
final state interaction of the active quark, gives rise to
the Sivers asymmetry in momentum space. We demonstrate 
this phenomenon explicitly in the scalar diquark model.
This result also illustrates the physics that underlies the 
correlation between the anomalous magnetic moment 
and the Sivers asymmetry  for a given quark flavor. 
\end{abstract}

\section{Introduction}
Recently, one gluon exchange in the final state interactions
(FSI) has been suggested \cite{BHS} as a mechanism for generating
a transverse single-spin asymmetry (SSA) in semi-inclusive
deep inelastic scattering processes. This FSI can be effectively
taken into account by introducing an appropriate 
Wilson line phase factor in the definition of the distribution 
functions of quarks in the nucleon \cite{sivers,collins,ji1,ji2}. 
The physical interpretation
of this Wilson line phase factor is that it describes the phase 
factor of the propagator for the active quark as it leaves the target.
Therefore, the Wilson line extends from the position of the quark
along a light-like future-oriented direction. This phase
factor is not invariant under time-reversal, which is the reason why
the Sivers asymmetry can be nonzero when the phase factor is included.

This mechanism has a nice physical interpretation in
transverse position space, where one can show that the Sivers
asymmetry arises from a left-right (relative to the nucleon spin)
asymmetry of the quark distribution in impact parameter space.
The Wilson line phase factor describes the effect of the
transverse component of the force that is acting
on the active quark after it is knocked out of the target. This
force is on average directed towards the center of the nucleon.
This {\it chromodynamic lensing} effect thus
translates the T-even transverse position space asymmetry
into a T-odd transverse momentum space asymmetry of the leading
quark \cite{mb1,mb2}, which is the Sivers asymmetry.

In Ref. \cite{BHS} a simple scalar diquark model was used
to demonstrate explicitly that the FSI can indeed give rise to
a leading-twist transverse SSA, which emerged from interference 
between spin dependent amplitudes with different 
nucleon spin states. 
In Refs. \cite{BHS,weak} it was observed that the
same overlap integrals between light-cone wavefunctions
that describe the anomalous magnetic moment contribution from
a given quark flavor also appear in the Sivers distribution for 
that quark flavor (with additional pieces in the integrand).
Since these integrals are the overlaps between light-cone wavefunctions
whose orbital angular momenta differ by $\Delta L^z = \pm 1$,
the orbital angular momentum of the quark inside the proton is
essential for the existence of the Sivers asymmetry.
In Refs. \cite{mb1,mb2}, the transverse distortion of impact 
parameter dependent parton distributions for transversely 
polarized targets in the
infinite momentum frame, was used to develop a physical 
explanation for the sign of the Sivers asymmetry.
In this paper, we diagonalize the relevant 
amplitudes in Ref. \cite{BHS} by transforming to an impact 
parameter space basis \cite{mb2}\footnote{For a general discussion
of parton distributions in impact parameter space and their 
connection to generalized parton distributions 
the reader is refered to Refs. \cite{me:1st,soper,diehl,gpde}.}.
This should not only yield a clear intuitive physical interpretation 
for the SSA derived in Ref. \cite{BHS}, but at the same time
provide an explicit example that illustrates a proposed
connection between generalized parton distributions and SSAs 
\cite{mb2}.

\bigskip\bigskip

\flushbottom

\section{GPDs in impact parameter space (scalar diquark
model)}
Light-cone wavefunctions provide a very convenient representation
of generalized parton distributions (GPDs) in terms of
overlap integrals \cite{DFJK,BDH}. 
%For purely transverse momentum transfer
%($\xi=0$), one finds
%\bea
%H(x,0,-\D2)&=& \sum_{n,\lambda_i} \int \prod_{i=1}^n
%\frac{dx_i d^2{\bf k}_{\perp i}}{16\pi^3}
%16\pi^3 \delta\left(1-\sum_{j=1}^n x_j\right)
%\delta^{(2)}\left(\sum_{j=1}^n{\bf k}_{\perp j}\right)\\
%& &\quad\quad\quad\quad\quad\times \delta(x-x_1)
%\tilde{\psi}_{(n)}^{\uparrow *}(x_i,{\bf k}_{\perp i}^\prime,\lambda_i)
%\tilde{\psi}_{(n)}^{\uparrow }(x_i,{\bf k}_{\perp i},\lambda_i)
%\nonumber \\
%\frac{\Delta^1-i\Delta^2}{2M}
%E(x,0,-\D2)&=& \sum_{n,\lambda_i} \int \prod_{i=1}^n
%\frac{dx_i d^2{\bf k}_{\perp i}}{16\pi^3}
%16\pi^3 \delta\left(1-\sum_{j=1}^n x_j\right)
%\delta^{(2)}\left(\sum_{j=1}^n{\bf k}_{\perp j}\right)\\
%& &\quad\quad\quad\quad\quad\times \delta(x-x_1)
%\tilde{\psi}_{(n)}^{\uparrow *}(x_i,{\bf k}_{\perp i}^\prime,
%\lambda_i)
%\tilde{\psi}_{(n)}^{\downarrow }(x_i,{\bf k}_{\perp i},\lambda_i),
%\nonumber
%\eea
%where the arguments of the final-state wavefunction are given by
%\bea
%\begin{array}{ll}
%{\bf k}_{\perp 1}^\prime = {\bf k}_{\perp 1} - (1-x_1)\DT
%&\mbox{for the struck quark}\\
%{\bf k}_{\perp i}^\prime = {\bf k}_{\perp i} +x_i\DT
%&\mbox{for the spectators $i=2,...,n$}\end{array}
%\eea
For example, in a scalar diquark model
in which a nucleon is modeled as a bound state of a quark and
a scalar diquark, one finds in the case
$\xi=0$ (purely transverse momentum transfer $
\Delta = p-p^\prime$)
\begin{eqnarray}
&&H(x,0,-\D2)
\nonumber\\
&&\qquad\ =\
\frac{1}{4\pi}\int {d^2{\bf k}_{\perp}\over (2\pi)^2}\
\left[ {\widetilde{\psi}}^{\uparrow\ *}_{+\frac{1}{2}}
(x,{\bf k}'_{\perp})
{\widetilde{\psi}}^{\uparrow}_{+\frac{1}{2}}(x,{\bf k}_{\perp})\ +\
{\widetilde{\psi}}^{\uparrow\ *}_{-\frac{1}{2}}(x,{\bf k}'_{\perp})
{\widetilde{\psi}}^{\uparrow}_{-\frac{1}{2}}(x,{\bf k}_{\perp})
\right]\ ,
\label{gf3} \\
&&{(\Delta^x-{i} \Delta^y)\over 2M}\ E(x,0,-\D2)
\nonumber\\
&&\qquad\ =\
\frac{1}{4\pi}\int {d^2{\bf k}_{\perp}\over (2\pi)^2}\
\left[ {\widetilde{\psi}}^{\uparrow\ *}_{+\frac{1}{2}}(x,{\bf k}'_{\perp})
{\widetilde{\psi}}^{\downarrow}_{+\frac{1}{2}}(x,{\bf k}_{\perp})\ +\
{\widetilde{\psi}}^{\uparrow\ *}_{-\frac{1}{2}}(x,{\bf k}'_{\perp})
{\widetilde{\psi}}^{\downarrow}_{-\frac{1}{2}}(x,{\bf k}_{\perp})\right]\ ,
\label{gf4}
\end{eqnarray}
where
\begin{equation}
{\bf k}'_{\perp}\ =\
{\bf k}_{\perp}-(1-x){\bf \Delta}_{\perp}\ .
\label{xprime}
\end{equation}
For higher Fock components, similar convolution formulas exist
\cite{DFJK,BDH}.

A well known feature of Fourier transforms is that they diagonalize
convolution integrals. For this purpose we switch to the
transverse position space representation of the light-cone
wavefunction
\bea
\psi^\lambda_s(x,\cT)\equiv 
\int\frac{d^2\kT}{(2\pi)^2} e^{i\kT\cT}
\tilde{\psi}^\lambda_s(x,\kT)
\eea
Note that the
transverse momentum $\kT$ in the two-particle Fock component
is Fourier conjugate to the distance 
$\cT\equiv
{\bf r}_{\perp 1} - {\bf r}_{\perp 2}$ between the active quark
and the spectator. However, since 
GPDs have a particularly simple form in impact parameter
representation, we change variables from 
%${\bf r_{\perp 1}} - {\bf r_{\perp 2}}$ 
$\cT$ to the impact parameter
$\bT$ (the distance
between the active quark and the center of longitudinal momentum),
which are related by
\bea
\bT= (1-x)\cT \ ,
\eea
yielding
\begin{eqnarray}
{\cal H}(x,{\bf b}_{\perp})&\equiv&
\int {d^2{\bf \Delta}_{\perp}\over (2\pi)^2}\ 
e^{i{\bf \Delta}_{\perp}\cdot{\bf b}_{\perp}}\
H(x,0,-\D2)
\label{b4}\\
&=&
{1 \over 4\pi}
\left[ {{\psi}}^{\uparrow\ *}_{+\frac{1}{2}}
(x,{\bf c}_{\perp})
{{\psi}}^{\uparrow}_{+\frac{1}{2}}(x,{\bf c}_{\perp})\ +\
{{\psi}}^{\uparrow\ *}_{-\frac{1}{2}}(x,{\bf c}_{\perp})
{{\psi}}^{\uparrow}_{-\frac{1}{2}}(x,{\bf c}_{\perp})\right]
\ {1\over (1-x)^2}\ .
\nonumber
\end{eqnarray}
Likewise one finds
\begin{eqnarray}
\frac{-i\frac{\partial}{\partial b^x}-\frac{\partial}{\partial b^y}
 }{2M} {\cal E}(x,\bT)&=&
{1 \over 4\pi}
\left[ {{\psi}}^{\uparrow\ *}_{+\frac{1}{2}}
(x,{\bf c}_{\perp})
{{\psi}}^{\downarrow}_{+\frac{1}{2}}(x,{\bf c}_{\perp})\ +\
{{\psi}}^{\uparrow\ *}_{-\frac{1}{2}}(x,{\bf c}_{\perp})
{{\psi}}^{\downarrow}_{-\frac{1}{2}}(x,{\bf c}_{\perp})\right]
\ {1\over (1-x)^2}\ ,\nonumber\\
& &
\label{b5}
\end{eqnarray}
where
\begin{eqnarray}
{\cal E}(x,{\bf b}_{\perp})&\equiv&
\int {d^2{\bf \Delta}_{\perp}\over (2\pi)^2}\
e^{i{\bf \Delta}_{\perp}\cdot{\bf b}_{\perp}}\
E(x,0,-\D2)\ .
\end{eqnarray}

The physical significance of ${\cal E}(x,{\bf b}_{\perp})$, i.e.,
the spin-flip distribution in impact parameter space
becomes clear when we consider a state that is
polarized in the $+\hat{y}$ direction (in the infinite momentum frame)
\be
\left|P^+,{\bf R}_{\perp}=\0T,+\hat{y}\right\rangle
\equiv{1\over {\sqrt{2}}}\left(\left|P^+,{\bf R}_{\perp}=\0T,
\uparrow \right\rangle +
i\left|P^+,{\bf R}_{\perp}=\0T,\downarrow \right\rangle\right).
\ee
For this state, the unpolarized quark distribution in impact 
parameter space reads
\begin{eqnarray}
q_{\hat{y}} (x,{\bf b}_{\perp})&\equiv&
\left\langle P^+,{\bf R}_{\perp}=\0T,+\hat{y}\right|
{\hat O}_q(x,{\bf b}_{\perp})
\left|P^+,{\bf R}_{\perp}=\0T,+\hat{y}\right\rangle
\nonumber\\
&=&\int {d^2{\bf \Delta}_{\perp}\over (2\pi)^2}\
e^{i{\bf \Delta}_{\perp}\cdot {\bf b}_{\perp}}\
\left[\ H(x,0,-\D2)+i{\Delta^x\over 2M}\
E(x,0,-\D2) \ \right]
\nonumber\\
&=&{\cal H}(x,{\bf b}_{\perp})+ {1\over 2M}{\partial\over\partial b^x}
{\cal E}(x,{\bf b}_{\perp})\ .
\label{b7}
\end{eqnarray}
%(In the second line of (\ref{b7}), the argument of the exponential
%has positive sign since
%${\bf P}_{\perp}={\bf P}'_{\perp}+{\bf \Delta}_{\perp}$ is used
%here.)

We will write in Section \ref{sec:X} 
on the interesting (and important) property
that the density in impact parameter space 
$q_{\hat{y}}(x,{\bf b}_{\perp})$ has the asymmetry along the
$\hat{x}$-direction in impact parameter space
when the proton spin is polarized along the
$\hat{y}$-direction as we see in (\ref{b7}), whereas there does not
exist such an asymmetry for the density in momentum space.

\section{Transverse distortion of the wavefunction}
In Ref. \cite{gpde} it was shown that if the spin-flip
generalized parton distribution $E_q(x,0,-\D2)$ is nonzero,
then the parton distribution of quarks with flavor $q$ is
distorted in the transverse plane when the target has a
transverse polarization. For a nucleon with spin pointing in the
positive $\hat{y}$-direction, one finds \cite{gpde}
\be
q_{\hat{y}}(x,\bT) = {\cal H}(x,\bT) + \frac{1}{2M}\frac{\partial}{
\partial b^x} {\cal E}(x,\bT),
\label{eq:11}
\ee
where ${\cal H}$ and ${\cal E}$ are Fourier transforms of
generalized parton distributions
\bea
{\cal H}_q(x,\bT)&\equiv& \int \frac{d^2\DT}{(2\pi)^2}
e^{i\DT\cdot \bT} H_q(x,0,-\D2)\nonumber\\
{\cal E}_q(x,\bT)&\equiv& \int \frac{d^2\DT}{(2\pi)^2}
e^{i\DT\cdot \bT}E_q(x,0,-\D2).
\eea
One of the things that are known about
${\cal E}_q$ is that its integral should give the contribution from 
flavor $q$ to the anomalous magnetic moment\footnote{The 
measured magnetic moment of the proton is obtained as a
superposition from all quark flavors
$\kappa_p = \sum_q e_q \kappa_q =1.79$.}
\be
\int dx \int d^2\bT {\cal E}_q(x,\bT)=\kappa_q.
\ee
Integrating $E(x,0,-\D2)$ over $x$ yields the Pauli form factor
$F_2(-\D2)$. Therefore, unless the $x$ dependence introduces 
a fluctuating sign into $E(x,0,-\D2)$, the sign of $F_2$ thus
determines the sign of $E(x,0,-\D2)$. Hence one expects
that $E(x,0,-\D2)$ is a smooth function of $\DT$ with a maximum
(or minimum if $\kappa_q<0$) at the origin. From experience with
Fourier transforms it is thus clear that ${\cal E}(x,\bT)$
also has a maximum (minimum) at the origin and is otherwise a
smooth function that will have the same sign as $\kappa_q$ for
all (or most) values of $\bT$. 

The $b^x$-derivative of a smooth positive function 
${\cal E}(x,{\bf b_\perp})$ with maximum at the
origin (we consider here the case $\kappa_q>0$)
is positive for negative $b^x$ and negative for positive
$b^x$. 
For valence quarks, ${\cal H}(x,\bT)$ is known to be positive.
Therefore, adding the $b^x$-derivative of ${\cal E}(x,\bT)$ to
${\cal H}(x,\bT)$ has the effect of shifting the
distribution towards negative $b^x$. For negative $\kappa_q$ the 
effect is reversed. This is why the sign of $\kappa_q$ determines the 
sign of the distortion of the quark distribution in impact parameter 
space. For a nucleon that is polarized in the $+\hat{y}$ direction the
distortion is towards negative $\hat{x}$ when $\kappa_q>0$ and 
towards positive $\hat{x}$ when $\kappa_q<0$.

Note that although the argument hinges somewhat on assumptions
about the shape of $E_q(x,0,\D2)$, these assumptions seem
to be satisfied for typical model ans\"atze for GPDs and
therefore the result is actually rather general.
In the specific example of the scalar diquark model, we can
calculate ${\cal E}_q(x,\bT)$ and verify that the model
satisfied the above assumptions.

Since the scalar diquark model provides us also with the
light-cone wavefunction for the `nucleon' there is actually
a more direct way to determine the sign of the distortion
in impact parameter space.
For this purpose we consider the wavefunction of a quark
that is polarized in the $+\hat{y}$-direction \cite{BD80,BHMS}
\bea
\label{eq:psiy}
\tilde{\psi}^{+\hat{y}}_{+\frac{1}{2}}(x,\kT)
&\equiv& \frac{1}{\sqrt{2}}\left[
\tilde{\psi}^{\uparrow}_{+\frac{1}{2}}(x,\kT)
+i\tilde{\psi}^{\downarrow}_{+\frac{1}{2}}(x,\kT)\right]
\\ &=& \frac{1}{\sqrt{2}}
\left[M+\frac{m}{x} + \frac{ik^x+k^y}{x}\right] 
\tilde{\phi}(x,\kT) \nonumber\\
\tilde{\psi}^{+\hat{y}}_{-\frac{1}{2}}(x,\kT)
&\equiv& \frac{1}{\sqrt{2}}\left[
\tilde{\psi}^{\uparrow}_{-\frac{1}{2}}(x,\kT)
+i\tilde{\psi}^{\downarrow}_{-\frac{1}{2}}(x,\kT)\right]
\label{eq:psiy2}
\\ &=& \frac{1}{\sqrt{2}}
\left[i\left(M+\frac{m}{x}\right) - \frac{k^x+ik^y}{x}\right] 
\tilde{\phi}(x,\kT) \nonumber .
\eea
Although this should be evident from time-reversal invariance,
we note that the naive (i.e. gauge noninvariant) unintegrated
momentum space distribution obtained from the wavefunction 
[described by Eqs. (\ref{eq:psiy},\ref{eq:psiy2})] 
squared is even in $\kT$. Indeed
\bea
{\widetilde{q}}_{\hat{y}}(x,\kT)=
{1\over 4\pi}\Bigl( \,
\left|\tilde{\psi}^{+\hat{y}}_{+\frac{1}{2}}(x,\kT)\right|^2
+\left|\tilde{\psi}^{+\hat{y}}_{-\frac{1}{2}}(x,\kT)\right|^2
\,\Bigr)
= {1\over 4\pi}\left[ \left(M+\frac{m}{x}\right)^2 +
\frac{{\bf k}_\perp^2}{x^2}\right]\tilde{\phi}^2,
\label{keven}
\eea
where we used the fact that $\tilde{\phi}$ is real.
(See Eq. (\ref{wfdenom}) in Appendix A.)

However, a transverse asymmetry in position space is not 
excluded by time-reversal invariance and Eqs. 
(\ref{eq:psiy},\ref{eq:psiy2}) do in fact correspond to a state with
an asymmetry in the $\hat{x}$-direction as we will now
demonstrate explicitly.
After performing a Fourier transformation to the transverse
relative position space coordinate $\cT$, we have
\be
\psi^{+\hat{y}}_{+\frac{1}{2}}(x,\cT)&\equiv& \int
\frac{d^2\kT}{(2\pi)^2}e^{i\cT\cdot \kT}
\tilde{\psi}^{+\hat{y}}_{+\frac{1}{2}}(x,\kT)
\nonumber\\
&=&\frac{1}{\sqrt{2}}
\left[M+\frac{m}{x} + \frac{1}{x} \frac{d}{dc^x} -
\frac{i}{x} \frac{d}{dc^y}\right]\phi(\cT)
\label{lcfw1}\\
\psi^{+\hat{y}}_{-\frac{1}{2}}(x,\cT)&\equiv& \int
\frac{d^2\kT}{(2\pi)^2}e^{i\cT\cdot \kT}
\tilde{\psi}^{+\hat{y}}_{-\frac{1}{2}}(x,\kT)
\nonumber\\
&=&\frac{1}{\sqrt{2}}\left[\frac{i}{x} \frac{d}{dc^x} -
\frac{1}{x} \frac{d}{dc^y} +i\left(M+\frac{m}{x} \right)\right]
\phi(\cT)
\label{lcfw2}
\ee
with
\bea
\phi (\cT) &\equiv&  \int \frac{d^2\kT}{(2\pi)^2}e^{i\cT\cdot \kT} 
\tilde{\phi} (\kT) \nonumber\\
&=& -g x\sqrt{1-x} \int\frac{d^2\kT}{(2\pi)^2}e^{i\cT\cdot \kT}  \frac{1}
{\kT^2 +B}
%\nonumber\\ &=& \frac{g}{4\pi} x\sqrt{1-x} \ln\left(\cT^2B\right), 
\nonumber\\ &=& 
-\, {g\over 2\pi}\, x\sqrt{1-x}\,\, K_0(|c_\perp| \sqrt{B})\ ,
\label{eq:phi}
\eea
and
\begin{equation}
B={x} (1-{x})(-M^2+{m^2\over {x}}+{{\lambda}^2\over 1-{x}})\ .
\label{b12}
\end{equation}
In the limit $|c_\perp| \sqrt{B} \to 0$, (\ref{eq:phi}) becomes
$({g}/{4\pi})\, x\sqrt{1-x}\, \ln\left(\cT^2B\right)$.
Using (\ref{lcfw1}) and (\ref{lcfw2}), we have
\begin{equation}
q_{\hat{y}}(x,\cT)=
{1\over 4\pi}\Bigl( \,
\left| \psi^{+\hat{y}}_{+\frac{1}{2}}(x,\cT)\right|^2
+ \left| \psi^{+\hat{y}}_{-\frac{1}{2}}(x,\cT)\right|^2
\,\Bigr)
\label{qyct}
\end{equation}
with
\begin{eqnarray}
\left| \psi^{+\hat{y}}_{+\frac{1}{2}}(x,\cT)\right|^2
= \left| \psi^{+\hat{y}}_{-\frac{1}{2}}(x,\cT)\right|^2
&=& \frac{1}{2}\left\{\left[ \left(M+\frac{m}{x} \right)\phi
+\frac{1}{x} \frac{d}{dc^x} \phi \right]^2
+ \frac{1}{x^2} \left(\frac{d}{dc^x} \phi\right)^2\right\}\nonumber\\
&=& \frac{1}{2}\left\{
\left(M+\frac{m}{x} \right)\phi^2 + \frac{1}{x^2}
\left[\left(\frac{d}{dc^x} \phi\right)^2+
\left(\frac{d}{dc^y} \phi\right)^2\right]\right\}
\nonumber\\
& &+ \frac{1}{x}  \left(M+\frac{m}{x} \right)\phi \frac{d}{dc^x}\phi
\ ,
\label{qycta}
\end{eqnarray}
since $\phi(\cT)$ is real.
In (\ref{qycta})
the last term is odd under $c^x \rightarrow -c^x$ and describes
the deformation of the target in impact parameter space
as predicted by Eq. (\ref{eq:11}):
\bea
\frac{1}{2M}\frac{\partial}{\partial b^x} {\cal E}(x,\bT)
= \frac{1}{4\pi} \ {1\over (1-x)^2}
\frac{2}{x}  \left(M+\frac{m}{x} \right)\phi \frac{d}{dc^x}\phi
\ .
\eea
 
Since $\phi(x,\cT)$ is a monotonically decreasing function of
${\bf c}_\perp^2$, this implies that
$\left| \psi^{+\hat{y}}_{\pm\frac{1}{2}}(x,\cT)\right|^2$
distorted towards negative $c^x$. For example, the mean $\perp$ 
coordinate for fixed $x$ yields
\bea
\label{bodd}
\langle c^x\rangle &\equiv& \int d^2\cT
\,\, {1\over 4\pi}
\left| \psi^{+\hat{y}}_{\pm\frac{1}{2}}(x,\cT)\right|^2
c^x
\\
&=& - \, {1\over 4\pi} \,
\frac{1}{2x}\left(M+\frac{m}{x}\right) \int d^2\cT \phi^2\
<\ 0 \ .
\nonumber
\eea

\section{SSA in impact parameter space (scalar diquark
model)}
The SSA for the scalar diquark model has been calculated
in Ref. \cite{BHS}. For a target that is polarized in the
$+\hat{y}$ direction, one finds for the $\perp$ momentum distribution
of the outgoing quarks\footnote{Note that there should be an
additional overall minus sign in front of Eq. (21) of Ref. \cite{BHS}.}
of quarks that carried
light-cone momentum fraction $x$
\begin{eqnarray}
&&{\cal P}^{\hat{y}}({x} , {\bf r}_{\perp})\ f_1({x} , {\bf r}_{\perp})
\nonumber\\
&=& C\ ({x} M+m)\ (1-{x})\ {1\over {\bf r}_{\perp}^2+B}\
\int {d^2{\bf k}_{\perp}\over (2\pi)^2}\
{1\over {\bf k}_{\perp}^2+B}\
{(k_\perp - r_\perp )^x\over ({\bf k}_{\perp}-{\bf r}_{\perp})^2
+{\lambda_g}^2}\ ,
\label{b11}
\end{eqnarray}
where $C=g^2e_1e_2/(2(2\pi)^3)$ and
$B$ is given in (\ref{b12}).
Here ${\cal P}^{\hat{y}}({x} , {\bf r}_{\perp})$ is the actual
spin asymmetry and $ f_1({x} , {\bf r}_{\perp})$ is the unpolarized
quark distribution so that the product ${\cal P}^{\hat{y}}f_1$ is
the spin-odd part of the outgoing quark momentum distribution.

In the following we are going to investigate within the context of
the scalar diquark model if this SSA can be related to GPDs
and the asymmetry of PDFs in impact parameter space, as has been
conjectured in Ref. \cite{mb2}.
Eq. (\ref{b11}) yields for the average transverse momentum
in the $\hat{x}$ direction
\begin{eqnarray}
\langle r^x\rangle &\equiv&\int d^2{\bf r}_{\perp}\ r^x
{\cal P}^{\hat{y}}({x} , {\bf r}_{\perp})\ f_1({x} , {\bf r}_{\perp})
\\
&=& C\
({x} M+m)\ (1-{x})
\int d^2{\bf r}_{\perp}\
\int {d^2{\bf k}_{\perp}\over (2\pi)^2}\
{r_\perp^x \over {\bf r}_{\perp}^2+B}\
{1\over {\bf k}_{\perp}^2+B}\
{ (k_\perp - r_\perp )^x
\over ({\bf k}_{\perp}-{\bf r}_{\perp})^2
+{\lambda_g}^2}
\nonumber
\end{eqnarray}
Using Eq. (\ref{eq:phi})
\begin{equation}
\frac{1}{{\bf k}_{\perp}^2+B} = -\frac{1}{g{x} \sqrt{1-{x}}}
\int {d^2 \cT^\prime} e^{i\kT\cdot\cT^\prime} \phi(\cT^\prime)
\end{equation}
and
\begin{equation}
\frac{1}{{\bf r}_{\perp}^2+B} = -\frac{1}{g{x} \sqrt{1-{x}}}
\int {d^2 \cT} e^{-i\rT\cdot\cT} \phi(\cT)
\end{equation}
as well as the relation (\ref{ap3}) given in Appendix B, 
from (\ref{b11}) we get
\begin{eqnarray}
\langle r^x\rangle
&=& 
\frac{e_1e_2}{2(2\pi)^3} \frac{\left({x} M+m\right)}{{x}^2}
\int d^2{\bf r}_{\perp}\
\int {d^2{\bf k}_{\perp}\over (2\pi)^2}\
\int {d^2 \cT} e^{-i\rT\cdot\cT} 
\int {d^2 \cT^\prime} e^{i\kT\cdot\cT^\prime} \nonumber\\
& &\times\
\phi(\cT^\prime)\left(-i\frac{\partial}{\partial c_x}\right)
\phi(\cT)
{ (k_\perp - r_\perp )^x
\over ({\bf k}_{\perp}-{\bf r}_{\perp})^2
+{\lambda_g}^2}
\nonumber\\
&=&
\frac{-ie_1e_2}{2(2\pi)} \frac{({x} M+m)}{{x}^2}
\int d^2\cT 
\phi(\cT)\frac{\partial}{\partial c_x}\phi(\cT)
\int {d^2{\bf k}_{\perp}\over (2\pi)^2}\ e^{i\kT\cdot\cT} 
{ k^x_\perp
\over {\bf k}_{\perp}^2
+{\lambda_g}^2}
\nonumber\\
&=& 
\int d^2{\bf b}_{\perp}\ {\bf I}_\perp^x(\cT) \frac{1}{2M} 
\frac{\partial}{\partial b^x} {\cal E}(x,\bT),
\label{b13}
\end{eqnarray}
where
%\bea
%f({\bf x}_{\perp})&=&
%\int {d{\bf k}_{\perp}\over (2\pi)^2}\ e^{i{\bf x}_{\perp}\cdot {\bf k}_{\perp}}
%\ {1\over {\bf k}_{\perp}^2+B}\ ,
%\label{b14}
%\eea
%and
\bea
{\bf I}_\perp^x({\bf c}_{\perp})&\equiv& -i\frac{e_1e_2}{2}
\int {d^2{\bf k}_{\perp}\over (2\pi)^2}\
e^{i\cT\cdot \kT }\
{ k_\perp^x 
\over {\bf k}_{\perp}^2
+{\lambda_g}^2}\ 
\label{b15}\\
&\stackrel{\lambda_g \rightarrow 0}{\longrightarrow}
& \frac{e_1e_2}{4\pi}\frac{ c_\perp^x}{{\bf c}_\perp^2 }
\end{eqnarray}
is the transverse impulse as a function of the $\perp$ coordinate
of the struck quark.

The physical interpretation as a `transverse impulse'  
becomes clear when one compares 
$\IT$ to the net transverse impulse $\int dt {\bf F}_\perp$ 
that one obtains when
one integrates the transverse component of the Coulomb force 
along a straight line from position $(c_\perp^x,c_\perp^y,0)$
to $(c_\perp^x,c_\perp^y,+\infty)$ along a straight line
(remember: for an ultrarelativistic particle, $z(t)=t$)
\be
\int_0^\infty dt\ {\bf F}_\perp\left(\cT,z(t)\right)
=\frac{e_1e_2}{4\pi}
\int_0^\infty dt \frac{\cT}{\left({\bf c}_\perp^2
+t^2\right)^{\frac{3}{2}}} = \frac{e_1e_2}{4\pi} 
\frac{\cT}{{\bf c}_\perp^2}
\ee

We note that Eq. (\ref{b13}) can be written as
\begin{eqnarray}
\langle r^x\rangle
&=&\int d^2{\bf b}_{\perp}\ {\bf I}_\perp^x(\cT) \
\frac{1}{2M}\frac{\partial}{\partial b^x} {\cal E}(x,\bT)
\nonumber\\
&=&\int d^2{\bf b}_{\perp}\ {\bf I}_\perp^x(\cT) \
q_{\hat{y}}^{\rm asym}(x,\bT)
\nonumber\\
&=&\int d^2{\bf b}_{\perp}\ {\bf I}_\perp^x(\cT) \
\Big[q_{\hat{y}}^{\rm sym}(x,\bT)
+q_{\hat{y}}^{\rm asym}(x,\bT)
\Big]
\nonumber\\
&=&
\int d^2{\bf b}_{\perp}\ {\bf I}_\perp^x(\cT) \
q_{\hat{y}}(x,\bT),
\label{rx2}
\end{eqnarray}
where $q_{\hat{y}}^{\rm sym}(x,\bT)$ and $q_{\hat{y}}^{\rm asym}(x,\bT)$
are symmetric and asymmetric about $b_{\perp}^x=0$, and they are given by
$q_{\hat{y}}^{\rm sym}(x,\bT)={\cal H}(x,\bT)$ and
$q_{\hat{y}}^{\rm asym}(x,\bT)=
\frac{1}{2M}\frac{\partial}{\partial b^x} {\cal E}(x,\bT)$.
This relation between GPDs and SSAs, which we now derived
explicitly for the scalar diquark model, is exactly of the
form ``$SSA\ = GPD\ *\ FSI$'' that was
proposed in Ref. \cite{mb2}.

We emphasize that $q_{\hat{y}}^{\rm asym}(x,\bT)$ is not zero in the
impact parameter space and this fact made the expressions given in
Eq. (\ref{b13}) and (\ref{rx2}) possible, however,
${\widetilde{q}}_{\hat{y}}^{\rm asym}(x,\kT)$
in the momentum space is identically zero: \label{sec:X}
\begin{equation}
{\widetilde{q}}_{\hat{y}}(x,\kT)={\widetilde{q}}_{\hat{y}}^{\rm sym}(x,\kT)\ ,
\label{qkt}
\end{equation}
as we demonstrated explicitly in Eq. (\ref{keven})
in the scalar diquark model
%explicitly
%from ${\widetilde{q}}_{\hat{y}}(x,\kT)={1\over {\sqrt{2}}}
%\Bigl( {\widetilde{\psi}}^{\uparrow}_s(x,\kT)+
%i{\widetilde{\psi}}^{\uparrow}_s(x,\kT){\Bigr)}^*
%{1\over {\sqrt{2}}}
%\Bigl( {\widetilde{\psi}}^{\uparrow}_s(x,\kT)+
%i{\widetilde{\psi}}^{\uparrow}_s(x,\kT){\Bigr)}$
which has the wavefunctions given in Eqs. (\ref{sn2}) and (\ref{sn2a}).
Even though there exists such a difference between the asymmetric 
parts of
$q_{\hat{y}}(x,\bT)$ and ${\widetilde{q}}_{\hat{y}}(x,\kT)$,
they satisfy the following relation which is
consequence of a general property of Fourier transforms:
\begin{equation}
\int d^2{\bf b}_{\perp}\ q_{\hat{y}}(x,\bT)=
\int \frac{d^2 \kT}{(2\pi)^2}\ 
{\widetilde{q}}_{\hat{y}}(x,\kT)\ ,
\label{qbtkt}
\end{equation}
i.e., the norm (integral of the absolute square of the function) 
is invariant under a Fourier transform.

Our calculations also illustrate explicitely a very important 
difference between impact parameter dependent parton distributions
$q_{\hat{y}}(x,\bT)$ and unintegrated parton distributions 
$\tilde{q}_{\hat{y}}(x,\kT)$ for a transversely polarized target:
Even for a transversely polarized target the unintegrated parton
distribution $\tilde{q}_{\hat{y}}(x,\kT)$ has to be symmetric
under $\kT \rightarrow -\kT$ because of parity and time-reversal
arguments. Roughly speaking, the reason is that given only the 
spin vector
${\vec S}$  and the momentum vector ${\vec P}$ of the target it
is not possible to construct a parity and time-reversal invariant
correlation between the target spin and the quark momentum
${\vec k}$ that is odd under ${\vec k}\rightarrow -{\vec k}$.
For example, although the term
${\vec k}\cdot \left({\vec S} \times {\vec P}\right)$ is invariant
under parity, it is odd under time-reversal and therefore 
cannot appear in the momentum distribution for the
quark\footnote{This argument is also the reason why a (time-reversal
invariance breaking) initial or final state interaction is crucial 
for the Sivers asymmetry in QCD.}.
However, the situation is different for the transverse position
$\bT$ where an asymmetry with respect to $\bT\rightarrow -\bT$ is allowed:
To see this we consider the term  
${\vec b}\cdot \left({\vec S} \times {\vec P}\right)$ which is clearly
invariant under both parity and time-reversal. The fact that
one can write down a term proportional to the spin that is odd in
${\vec b}$ and consistent with the symmetries of QCD means that a
transverse (relative to ${\vec P}$ and ${\vec S}$) asymmetry in
the position space distribution is allowed.

The explicit model calculations in the scalar diquark model
confirm these general considerations since they yield
$\tilde{q}_{\hat{y}}(x,-\kT)=\tilde{q}_{\hat{y}}(x,\kT)$
[Eq. (\ref{keven})], while
${q}_{\hat{y}}(x,-\bT) \neq {q}_{\hat{y}}(x,\bT)$ 
[Eq. (\ref{bodd})].

We should also point out that Eq. (\ref{rx2}) is consistent
with expressions that have been written down previously
for the mean transverse momentum arising from the final state 
interactions \cite{QS,BMP}. What is new is that we are evaluating these
expressions in impact parameter space and for a specific model for
the final state interactions.

\section{Summary}
We have analyzed the single-spin asymmetry in
semi-inclusive deep inelastic scattering (DIS) in the scalar diquark
model. In the impact parameter representation,
the SSA emerges as a correlation between the
distribution of partons in the transverse plane
and the transverse impulse, which the quark
being ejected from a certain transverse position
has acquired as a result of the final state 
interactions.

The scalar diquark model is a model where both the
light-cone wavefunctions of the quarks as well as 
the final state interactions are constructed 
perturbatively and therefore the constructed
amplitudes still have all Lorentz symmetries as well
as gauge invariance built in. Because of all these
features, the physics of the Wilson line phase factor
that describes the final state interaction in
semi-inclusive DIS should be correctly represented
in this model.
In this model we obtained in the impact parameter space the formulas
(\ref{b13}) and (\ref{rx2}) which show explicitly the connection
between the SSA and ${\cal E}(x,\bT)$ which is the Fourier transform
of the generalized parton distribution $E_q(x,0,-\D2)$.
These formulas are also useful for understanding the role of the
orbital angular momentum in the Sivers asymmetry, since $E_q(x,0,-\D2)$
is given by the overlap integrals between light-cone wavefunctions
whose orbital angular momenta differ by $\Delta L^z = \pm 1$
\cite{BHS,BDH,BHMS}.

Without the final state interaction, one should
not find a transverse momentum asymmetry because of
time-reversal invariance. However, time-reversal
invariance does not exclude a transverse position
space asymmetry when the nucleon is polarized transversely. 
Our explicit calculations within the context of the
scalar diquark model confirm these general predictions.
We also find that the transverse position space asymmetry 
is described by the 
Fourier transform of the generalized parton distribution
$E(x,0,-\D2)$ as one would expect \cite{gpde}.
The transverse single-spin asymmetry in the model
is obtained by convoluting the final state 
interaction kernel with the momentum space light-cone
wavefunctions of the `quarks' in the `nucleon'.
Upon Fourier transforming the SSA to impact 
parameter space, one finds that the average SSA
has a probabilistic interpretation in impact parameter
space in the sense that
the asymmetry can be obtained by averaging the
transverse impulse for each point in impact
parameter space with the probability density to
find a quark at that impact parameter. On the one hand, this
result illustrates clearly the physical mechanism for the results
in Ref. \cite{BHS}, and on the other hand it
provides a specific example that
confirms the general mechanism for SSA, that was proposed 
in Refs. \cite{mb1,mb2}. 

\appendix
\section{Scalar diquark model}
\label{sec:diquark}
One of the virtues of light-cone wavefunctions is that many 
form factors and transition matrix elements have very simple
representations as overlap integrals involving these wavefunctions
\cite{LB,BPP}.
Of course, in order to apply these overlap integrals, one 
needs to know the light-cone wavefunctions for each Fock component.
Since we do not know the light-cone wavefunctions for the nucleon
in QCD, but would still like to use them to illustrate the connection
between transverse single-spin asymmetries and impact parameter
space distributions, we take them from a simple toy model
(a scalar diquark model).

%The contribution from the two particle Fock component to the
%generalized parton distributions $H$ and $E$ for $\xi=0$
%reads
%\bea
%H(x,0,-\D2) &=& \int \frac{d^2\kT}{16\pi^3}
%\left[\tilde{\psi}^{\uparrow*}_{+\frac{1}{2}}(x,\kTp)
%\tilde{\psi}^\uparrow_{+\frac{1}{2}}(x,\kT)
%+ \tilde{\psi}^{\uparrow*}_{-\frac{1}{2}}(x,\kTp)
%\tilde{\psi}^\uparrow_{-\frac{1}{2}}(x,\kT)\right]
%\\
%E(x,0,-\D2) &=& \int \frac{d^2\kT}{16\pi^3}
%\left[\tilde{\psi}^{\uparrow*}_{+\frac{1}{2}}(x,\kTp)
%\tilde{\psi}^\uparrow_{-\frac{1}{2}}(x,\kT)
%+ \tilde{\psi}^{\uparrow*}_{-\frac{1}{2}}(x,\kTp)
%\tilde{\psi}^\uparrow_{+\frac{1}{2}}(x,\kT)\right],
%\eea
%where
%\bea \kTp=\kT-(1-x)\DT .\eea

In the scalar diquark model, the light-cone
wavefunction for the two-particle Fock component
is obtained by calculating the `splitting' of a fermion
into another fermion plus a scalar perturbatively. This yields
the two-particle Fock component wavefunctions
of the $J^z = + {1\over 2}$ state \cite{BD80,BHMS}:
\begin{eqnarray}
\!\!\!&&\left|\Psi^{\uparrow}_{\rm two \ particle}(P^+, {\bf P}_\perp = 
{\bf 0}_\perp)\right>
\label{sn1} \\
\!\!\!&=& \int\frac{d^2 {\bf k}_{\perp} dx
}{{\sqrt{x(1-x)}}16 \pi^3} \left[ \ 
{\widetilde{\psi}}^{\uparrow}_{+\frac{1}{2}}
(x,{\bf k}_{\perp})\, \left| +\frac{1}{2}\, ;\,\, xP^+\, ,\,\,
{\bf k}_{\perp} \right>  %\\ && \qquad 
+
{\widetilde{\psi}}^{\uparrow}_{-\frac{1}{2}} (x,{\bf k}_{\perp})\,
\left| -\frac{1}{2}\, ;\,\, xP^+\, ,\,\, {\bf k}_{\perp} \right>\
\right]\ , \nonumber
\end{eqnarray}
where
\begin{equation}
\left
\{ \begin{array}{l}
{\widetilde{\psi}}^{\uparrow}_{+\frac{1}{2}}
(x,{\bf k}_{\perp})=(M+\frac{m}{x})\,
{\widetilde{\phi}} \ ,\\
{\widetilde{\psi}}^{\uparrow}_{-\frac{1}{2}} (x,{\bf k}_{\perp})=
-\frac{(k^x+ i k^y)}{x }\,
{\widetilde{\phi}} \ ,
\end{array}
\right.
\label{sn2}
\end{equation}
and
%The scalar part of the wavefunction $\phi$ depends on the 
%dynamics. In the perturbative theory it is simply
\begin{equation}
{\widetilde{\phi}}
={\widetilde{\phi}}(x,{\bf k}_{\perp})=
{g\over \sqrt{1-x}}\ \frac{1}{M^2-{{\bf k}_{\perp}^2
+m^2\over x}-{{\bf k}_{\perp}^2+\lambda^2\over 1-x}}\ .
\label{wfdenom}
\end{equation}
In general one normalizes the Fock state to unit probability.

Similarly, the $J^z = - {1\over 2}$ two-particle Fock state has components:
\begin{equation}
\left
\{ \begin{array}{l}
{\widetilde{\psi}}^{\downarrow}_{+\frac{1}{2}} (x,{\bf k}_{\perp})=
\frac{(k^x- i k^y)}{x }\,
{\widetilde{\phi}} \ ,\\
{\widetilde{\psi}}^{\downarrow}_{-\frac{1}{2}}
(x,{\bf k}_{\perp})=(M+\frac{m}{x})\,
{\widetilde{\phi}} \ .
\end{array}
\right.
\label{sn2a}
\end{equation}
The spin-flip amplitudes in (\ref{sn2}) and (\ref{sn2a}) have
orbital angular momentum projection $l^z =+1$ and $-1$
respectively.  The $\kT$ dependence in the numerators of the 
wavefunctions is
characteristic of the orbital angular momentum, and holds for both
perturbative and non-perturbative couplings.

\section{Useful relations for Fourier transformations}
\label{sec:fourier}
The fact that certain amplitudes that are convolutions in
momentum space become diagonal in position space can be
easily understood on the basis of some elementary theorems
about convolutions and Fourier transforms. For example, if
\bea
f(\kT)&=& \int {d^2\bT} e^{-i\kT\cdot\bT}
\tilde{f}(\bT) \nonumber\\
g(\kT)&=&\int {d^2\bT} e^{-i\kT\cdot\bT}
\tilde{g}(\bT) 
\eea 
then the ``form factor'' 
\bea
F(\qT)\equiv \int \frac{d^2\kT}{(2\pi)^2} f^*(\kT)g(\kT+\qT)
\eea
becomes diagonal in Fourier space
\bea
\int \frac{d^2 \qT}{(2\pi)^2} e^{i\qT\cdot\bT} F(\qT) = 
\tilde{f}^*(\bT) \tilde{g}(\bT).
\eea
This well-known result forms the basis for the interpretation
of non-relativistic form factors as charge distributions
in position space.

Similarly, the convolution describing the FSI in the amplitude has 
the form
\bea
A(\kT)= \int \frac{d^2 \qT}{(2\pi)^2} K(\kT-\qT) g(\qT)
\eea
and thus becomes diagonal in position space
\bea
\tilde{A}(\bT)\equiv \int \frac{d^2 \kT}{(2\pi)^2} e^{-i\kT\cdot\bT}
A(\kT) = \tilde{K}(\bT) \tilde{g}(\bT), \eea 
where 
\bea
\tilde{K}(\bT)\equiv \int \frac{d^2 \qT}{(2\pi)^2} K(\qT) e^{-i\qT\cdot\bT} 
\eea
is the Fourier transform of the kernel.

Finally, the mean SSA in momentum space 
contains a double convolution:
\bea
\left\langle \kT \right\rangle &=&
\int\frac{d^2 \kT}{(2\pi)^2} \int \frac{d^2 \qT}{(2\pi)^2}
f^*(\qT) K(\kT-\qT)g(\qT) \nonumber\\
&=& \int {d^2 \bT} \tilde{f}^*(\bT)
\tilde{K}(\bT) \tilde{g}(\bT).
\label{ap3}
\eea 

Of course, for practical purposes it is often more convenient
to stay in momentum space since experiments measure momenta and 
Fourier transforms are numerically awkward.
However, for the purpose of physical interpretation, the position
space expressions are useful since they are diagonal and therefore
allow a probabilistic interpretation.

%\begin{eqnarray}
%f(x)&=&\int {dk\over 2\pi}\ e^{ikx}\ {\widetilde{f}}(k)\ ,
%\label{ap1}\\
%{\widetilde{f}}(k)&=&\int dx\ e^{-ikx}\ f(x)\ .
%\label{ap1a}
%\end{eqnarray}
%
%\begin{equation}
%\int {dk'\over 2\pi}\ {\widetilde{f}}^*_1(k'-k)\
%{\widetilde{f}}_2(k')\ =\
%\int dx\ e^{-ikx}\ f^*_1(x)\ f_2(x)\ .
%\label{ap2}
%\end{equation}
%
%\begin{equation}
%\int {dk\over 2\pi}\ \int {dk'\over 2\pi}\
%{\widetilde{f}}^*_1(k)\ 
%{\widetilde{g}}(k-k')\ {\widetilde{f}}_2(k') \ =\
%\int dx\ f^*_1(x)\ g(x)\ f_2(x)\ .
%\label{ap3}
%\end{equation}

\section*{Acknowledgements}
We thank Daniel Boer, Stan Brodsky, and Markus Diehl
for very helpful comments.
M.B. was supported by the DOE under grant number DE-FG03-95ER40965.

\end{document}